\documentclass[10pt,amsmath]{revtex4}

\usepackage{graphicx}

\begin{document}

\makeatletter 
\def\@cite#1#2{\textsuperscript{{#1\if@tempswa}}}
\makeatother

\makeatletter
\def\@biblabel#1{#1.}
\makeatother

\title{Hyper-parallel photonic quantum computation with  coupled quantum dots\footnote{Published in  Sci. Rep. \textbf{4}, 4623  (2014).}}

\author{Bao-Cang Ren$^\dag $ and  Fu-Guo Deng\footnote{These authors
contributed equally to this work.}\footnote{   Correspondence and
requests for materials should be addressed to F. -G. Deng
(fgdeng@bnu.edu.cn).}}


\address{Department of Physics, Applied Optics Beijing Area Major Laboratory, Beijing Normal University, Beijing 100875, China}

\date{\today }

\begin{abstract}
It is well known that a parallel quantum computer is more powerful
than a classical one.  So far, there are some important works about
the construction of universal quantum logic gates,  the key elements
in quantum computation.  However, they are focused on operating on
one degree of freedom (DOF) of quantum systems. Here, we investigate
the possibility of achieving scalable hyper-parallel quantum
computation based on two DOFs of photon systems. We construct a
deterministic hyper-controlled-not (hyper-CNOT) gate operating on
both the spatial-mode and the polarization DOFs of a two-photon
system simultaneously, by exploiting the giant optical circular
birefringence induced by quantum-dot spins in double-sided optical
microcavities as a result of cavity quantum electrodynamics (QED).
This hyper-CNOT gate is implemented by manipulating the four qubits
in the two DOFs of a two-photon system without auxiliary spatial
modes or polarization modes. It reduces the operation time and the
resources consumed in quantum information processing, and it is more
robust against the photonic dissipation noise, compared with the
integration of several cascaded CNOT gates in one DOF.
\end{abstract}

\maketitle

Quantum information processing, based on quantum mechanics theory,
has largely improved the methods of dealing and transmitting
information in quantum computation\cite{QC} and quantum
communication\cite{rmp}. Parallel quantum computation provides a
novel way to precisely control and manipulate the states of quantum
systems, which is much faster than the conventional computation in
principle. Quantum logic gate is a key element in quantum
computation. It  is very important  to find an effective physical
realization of quantum logic gates. Many proposals  have been
proposed to implement controlled-not (CNOT) gates or controlled
phase-flip (CPF) gates both in theory and in experiment by using
various quantum systems, such as ion trap\cite{NTI},  free
electron\cite{FE}, cavity-QED system\cite{CQED,CQED2}, nuclear
magnetic resonance\cite{NMR,NMR2,NMR3}, quantum dot\cite{QDS},
superconducting charge qubits\cite{SCCQ}, and polarization states of
single photons\cite{linear2,KLM,nonlinear,QED1,QED2}. In 2001,
Knill, Laflamme, and Milburn\cite{KLM} showed that a CNOT gate on
photonic qubits with the maximal success probability of 3/4 could be
constructed, resorting to only single-photon sources, single-photon
detectors, and linear optical elements. In contrast, a deterministic
photonic CNOT gate can be constructed with nonlinear optics. In
2004, Nemoto and Munro\cite{nonlinear} constructed a nearly
deterministic CNOT gate with weak cross-Kerr nonlinearity, ancilla
photons, and feed-forward operations. The key element of this
proposal is  the quantum nondemolition detector using the nonlinear
optics of weak cross-Kerr media. At present, a giant Kerr
nonlinearity is still a challenge, even with electromagnetically
induced transparency\cite{nonlinear2}, because the initial  phase
shift achieved at the single-photon level\cite{nonlinear3} is only
in the order of 10$^{-5}$. Moreover, it is not clear whether these
nonlinearities are sufficient for the natural implementation of
single-photon qubit gates or not\cite{kerr}. Cavity  QED is a
promising candidate for obtaining large nonlinear phase shifts with
the dipole emitter resonant with the cavity mode\cite{QED}. In 2004,
Duan and Kimble\cite{QED1} proposed a scheme to implement a CPF gate
for photon systems with the nonlinear phase shifts caused by a
single atom trapped in an optical cavity. In 2010, Koshino et
al.\cite{QED2} presented a method to implement a deterministic
photon-photon $\sqrt{SWAP}$ gate using a three-level system embedded
in an optical cavity. The nonlinear approaches using multimode
optical parametric oscillator\cite{OPO} and coherent photon
conversion\cite{CPC} can also be used to implement deterministic
quantum computing processes.

In order to implement quantum computer, there are some  requirements
for physical systems: efficient manipulation, reading out the state
of an individual qubit, strong coupling between qubits, weak
coupling to environment, and scalability. Recently, a solid-state
quantum system based on an electron spin in a quantum dot (QD)
inside a microcavity (QD-cavity) has attracted much attention with
its good optical property and  scalability. In 2008, Hu et
al.\cite{QD1, QD2} proposed a quantum nondemolition method using the
giant optical circular birefringence of a one-side QD-cavity system,
and they pointed out that the interaction of circularly polarized
lights and double-sided QD-cavity system can be used to construct
entanglement beam splitter (EBS)\cite{QD3} in 2009. With the optical
property of QD-cavity systems, many quantum information processes
can be implemented, such as entanglement generation and Bell-state
analysis\cite{QD1,QD2,QD3,QD4}, quantum
gates\cite{QD6,QD5,QDWei1,QDWei2,QD7}, hyper-entangled Bell-state
analysis\cite{HBSA,HBSA1}, and quantum repeater\cite{repeaterwang}.
In 2010, Bonato et al.\cite{QD6} constructed a hybrid  quantum CNOT
gate on an electron-spin qubit and a photonic qubit using the EBS
effect of a double-sided QD-cavity system.

It is well known that any universal quantum computation can be
realized with two-qubit gates assisted by one-qubit rotations in
quantum network theory\cite{Unigate}. In practice, there are many
obstacles required to be overcome both in theory and in technology,
such as the sheer number of element gates in quantum logic circuits,
the implementation of element gates, and so on. So far, there are
many important works on the construction of CNOT gates and CPF gates
operating on one degree of freedom (DOF) of quantum systems (such as
the polarization DOF of photon systems or the spin DOF of electron
systems). Usually, the spatial-mode (polarization) DOF of photon
systems is used to assist the construction of quantum logic gates
that operate on the polarization (spatial-mode) DOF, in which the
spatial-mode (polarization) DOF is consumed after the operation. In
principle, both the spatial-mode and the polarization DOFs of photon
systems can be used as the qubits for universal quantum gates on two
DOFs of quantum systems. In this paper, we investigate the
possibility of achieving scalable hyper-parallel quantum computation
based on two DOFs of photon systems without using the auxiliary
spatial modes or polarization modes, which can reduce the resources
required for quantum logic circuits (or storage process) and depress
the photonic dissipation in quantum information processing, compared
with the integration of several cascaded CNOT gates in one DOF. We
construct a deterministic hyper-controlled-not (hyper-CNOT) gate
operating on both the spatial-mode and the polarization DOFs of a
photon pair simultaneously, resorting to the giant optical circular
birefringence induced by quantum-dot spins in double-sided optical
microcavities as a result of cavity QED. This hyper-CNOT gate is
useful for the quantum information protocols with multiqubit
systems, and it decreases the interaction times between the photons
required in quantum information processing.\\

\bigskip

{\large \textbf{Results}}

\begin{figure}[htbp]             
\centering\includegraphics[width=7.2 cm]{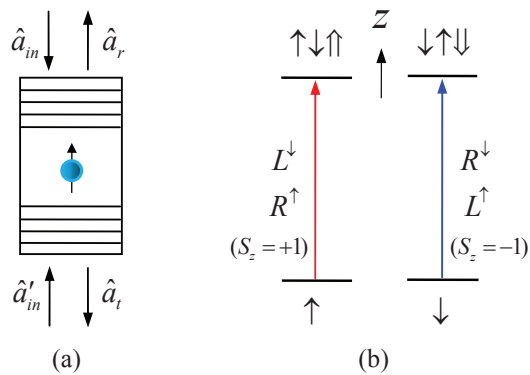} \caption{  (a) A
schematic diagram for a singly charged QD inside a double-sided
micropillar microcavity interacting with circularly polarized
lights. (b) The $X^-$ spin-dependent optical transition rules due to
the Pauli's exclusion principle. $L^\uparrow$ ($L^\downarrow$) and
$R^\uparrow$ ($R^\downarrow$) represent the left and the right
circularly polarized lights whose input directions are parallel
(antiparallel) with z direction, respectively. $\uparrow$ and
$\downarrow$ represent the spins $+\frac{1}{2}$ and $-\frac{1}{2}$
of the excess electron in QD, respectively.
$\uparrow\downarrow\Uparrow$ and $\downarrow\uparrow\Downarrow$
represent the  spin states of the negatively charged exciton $X^-$.
\label{figure1}}
\end{figure}

\textbf{Quantum-dot-cavity system.} Let us consider a singly charged
QD [e.g. self-assembled In(Ga)As QD or GaAs interfacial QD] confined
in a double-sided optical resonant microcavity. The QD is located at
the center of a double-sided cavity, shown in Fig.\ref{figure1}(a).
The two distributed Bragg reflectors are made partially reflective
and low loss for on-resonance transmission, and the double-sided
cavity is a two-mode one which supports both polarization
modes\cite{cavity}. With an excess electron injected into QD, the
singly charged QD shows the optical resonance of a negatively
charged exciton $X^-$ that consists of two electrons bound to one
hole\cite{QD9}. According to Pauli's exclusion principle, $X^-$
shows spin-dependent transitions interacting with circularly
polarized lights\cite{QD10}, shown in Fig.\ref{figure1}(b). The
left-circularly-polarized light $|L\rangle$ ($S_z=+1$) couples the
excess electron-spin state $|\uparrow\rangle$ to create $X^-$ in the
state $|\uparrow\downarrow\Uparrow\rangle$ with two electron spins
antiparallel, while the right-circularly-polarized light $|R\rangle$
($S_z=-1$) couples the excess electron-spin state
$|\downarrow\rangle$ to create $X^-$ in the state
$|\downarrow\uparrow\Downarrow\rangle$. Here $|\Uparrow\rangle$ and
$|\Downarrow\rangle$ represent the heavy-hole spin states
$|+\frac{3}{2}\rangle$ and $|-\frac{3}{2}\rangle$, respectively.

The input-output relation of this double-sided QD-cavity system can
be calculated from the Heisenberg equations of motion for the cavity
field operator $\hat{a}$ and $X^-$ dipole operator $\hat{\sigma}_-$
in the interaction picture\cite{QD11},
\begin{eqnarray}                           \label{eq.1}
\frac{d\hat{a}}{dt}&=&-[i(\omega_c-\omega)+\kappa+\frac{\kappa_s}{2}]\hat{a}-g\hat{\sigma}_{-} - \sqrt{\kappa}\,\hat{a}_{in}
-\sqrt{\kappa}\,\hat{a}'_{in}, \nonumber\\
\frac{d\hat{\sigma}_-}{dt}&=&-[i(\omega_{X^-}-\omega)+\frac{\gamma}{2}]\hat{\sigma}_--g\hat{\sigma}_z\hat{a},\nonumber\\
\hat{a}_{r}&=&\hat{a}_{in}+\sqrt{\kappa}\,\hat{a},\nonumber\\
\hat{a}_{t}&=&\hat{a}'_{in}+\sqrt{\kappa}\,\hat{a},
\end{eqnarray}
where  $\omega_{X^-}$, $\omega$, and $\omega_c$ are  the frequencies
of the $X^-$ transition, the input photon, and the cavity mode,
respectively. $g$ is the $X^-$-cavity coupling strength. $\gamma/2$,
$\kappa$, and  $\kappa_s/2$ are the decay rates of $X^-$, the cavity
field mode, and the cavity side leakage mode, respectively.
$\hat{a}_{in}$, $\hat{a}'_{in}$, and $\hat{a}_{r}$, $\hat{a}_{t}$
are the input and the output field operators, respectively.

In the approximation of a weak excitation condition with $X^-$
staying in the ground state at most time and
$\langle\sigma_z\rangle=-1$, the reflection and the transmission
coefficients of the $X^-$-cavity system can be expressed
as\cite{QD3},
\begin{eqnarray}                           \label{eq.2}
r(\omega)&=&1+t(\omega),\nonumber\\
t(\omega)&=&\frac{-\kappa[i(\omega_{X^-}-\omega)+\frac{\gamma}{2}]}{[i(\omega_{X^-}-\omega)
+\frac{\gamma}{2}][i(\omega_c-\omega)+\kappa+\frac{\kappa_s}{2}]+g^2}.\;\;\;\;
\end{eqnarray}
Considering the resonant condition  $\omega_c=\omega_{X^-}=\omega_0$
and the coupling strength $g=0$, the reflection and the transmission
coefficients $r_0(\omega)$ and $t_0(\omega)$ of a cold cavity (the
QD is uncoupled to the cavity) are obtained as\cite{QD3}
\begin{eqnarray}                           \label{eq.3}
r_0(\omega)&=&\frac{i(\omega_0-\omega)+\frac{\kappa_s}{2}}{i(\omega_0-\omega)+\kappa+\frac{\kappa_s}{2}},\nonumber\\
t_0(\omega)&=&\frac{-\kappa}{i(\omega_0-\omega)+\kappa+\frac{\kappa_s}{2}}.
\end{eqnarray}

For the input left-circularly-polarized light $|L\rangle$
($S_z=+1$), the QD with the electron spin state $|\uparrow\rangle$
is coupled to the cavity (a hot cavity) with  the reflection
coefficient $|r(\omega)|$ and the transmission coefficient
$|t(\omega)|$, whereas the QD with the electron spin state
$|\downarrow\rangle$ is uncoupled to the cavity (a cold cavity) with
the reflection coefficient $|r_0(\omega)|$ and the transmission
coefficient $|t_0(\omega)|$. Conversely, for the input
right-circularly-polarized  light $|R\rangle$ ($S_z=-1$), the QD
with the electron spin state $|\downarrow\rangle$ is coupled to the
cavity, whereas the QD with the electron spin state
$|\uparrow\rangle$ is uncoupled to the cavity. In the strong
coupling regime ($g>(\kappa,\gamma)$) with the resonant condition,
if we adjust the frequencies as $\omega\approx\omega_0$ and neglect
the cavity side leakage, we get $|r(\omega)|\rightarrow1$,
$|r_0(\omega)|\rightarrow0$ and $|t(\omega)|\rightarrow0$,
$|t_0(\omega)|\rightarrow1$. That is, the reflection  and the
transmission operators can be written as\cite{QD3}
\begin{eqnarray}                           \label{eq.4}
\hat{r}(\omega)&\simeq& r(\omega)(|R\rangle\langle
R|\;|\downarrow\rangle\langle\downarrow|+|L\rangle\langle
L|\;|\uparrow\rangle\langle\uparrow|),\nonumber\\
\hat{t}(\omega)&\simeq& t_0(\omega)(|R\rangle\langle
R|\;|\uparrow\rangle\langle\uparrow|+|L\rangle\langle
L|\;|\downarrow\rangle\langle\downarrow|).\;\;\;\;
\end{eqnarray}

Photonic circular polarization is usually dependent on the direction
of propagation, so the photon in the state $|R^\uparrow\rangle$ or
$|L^\downarrow\rangle$ has the spin $S_z=+1$, as shown in
Fig.\ref{figure1}(b). For the electron spin state
$|\uparrow\rangle$, there is the dipole interaction between the
electron spin and the photon with the photon state reflected to be
$|L^\downarrow\rangle$ or $|R^\uparrow\rangle$ by the cavity. For
the electron spin state $|\downarrow\rangle$, the input photon is
transmitted through cavity with a phase shift relative to the
reflected photon, that is\cite{QD6},
\begin{eqnarray}                           \label{eq.5}
&&\!\!\!\!|R^\uparrow, i_2, \uparrow\rangle \rightarrow
|L^\downarrow, i_2, \uparrow\rangle,\;\;\;\;\;\;\;\;\;\;\;\;\;\,
|L^\downarrow, i_1, \uparrow\rangle \rightarrow |R^\uparrow, i_1, \uparrow\rangle,\nonumber\\
&&\!\!\!\!|R^\uparrow, i_2, \downarrow\rangle \rightarrow
-|R^\uparrow, i_1\downarrow\rangle,\;\;\;\;\;\;\;\;\;\;\;
|L^\downarrow, i_1,\downarrow\rangle \rightarrow -|L^\downarrow,
i_2, \downarrow\rangle.\;\;\;\;\;\;\label{eq5}
\end{eqnarray}
In the same way, the reflection and the transmission rules of the
photon spin states $|R^\downarrow\rangle$ and $|L^\uparrow\rangle$
($S_z=-1$) can be obtained as
\begin{eqnarray}                           \label{eq.6}
&&\!\!\!\!|R^\downarrow, i_1, \uparrow\rangle \rightarrow
-|R^\downarrow, i_2, \uparrow\rangle,\;\;\;\;\;\;\;\;\;\;\,
|L^\uparrow, i_2, \uparrow\rangle \rightarrow -|L^\uparrow, i_1, \uparrow\rangle,\nonumber\\
&&\!\!\!\!|R^\downarrow, i_1, \downarrow\rangle \rightarrow
|L^\uparrow, i_1, \downarrow\rangle,\;\;\;\;\;\;\;\;\;\;\;\;\;\;
|L^\uparrow, i_2, \downarrow\rangle \rightarrow |R^\downarrow, i_2,
\downarrow\rangle.\;\;\;\;\;\;\;\;\label{eq6}
\end{eqnarray}
Here $i_1$ and $i_2$ ($i=a, b$) are the two spatial modes of the
photon $i$ (shown in Fig.2).

\begin{figure}[!h]
\centering\includegraphics[width=7.3 cm,angle=0]{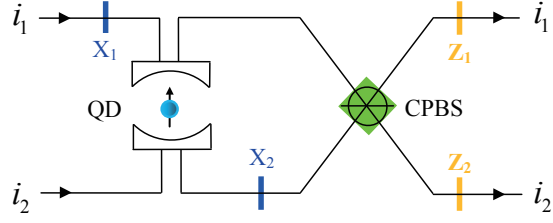}
\caption{Schematic diagram for a spatial-CNOT gate operating on the
spatial-mode DOF of a two-photon system.  $X_i$ ($i=1,2$) represents
a half-wave plate which is used to perform a polarization bit-flip
operation $X=|R\rangle\langle L|+|L\rangle\langle R|$ on the photon
passing through it. CPBS represents a polarizing beam splitter in
the circular basis, which transmits the photon in the polarization
$\vert R\rangle$ and reflects the photon in the polarization $\vert
L\rangle$, respectively. Z$_i$ ($i=1,2$) represents a half-wave
plate which is used to perform a polarization phase-flip operation
$Z=-|R\rangle\langle R|+|L\rangle\langle L|$ on a photon.
\label{figure2}}
\end{figure}

\bigskip

\textbf{Spatial-CNOT gate on a two-photon system.} The principle of
our spatial-CNOT gate is shown in Fig.\ref{figure2}. It is
constructed with the reflection and the transmission rules of
circularly polarized lights interacting with a QD-cavity system.
Suppose that the initial states of the electron spin in QD and the
photons $a$ and $b$ are
$\frac{1}{\sqrt{2}}(|\uparrow\rangle+|\downarrow\rangle)_e$,
$(\alpha_1|R\rangle+\alpha_2|L\rangle)_a(\gamma_1|a_1\rangle+\gamma_2|a_2\rangle)$,
and
$(\beta_1|R\rangle+\beta_2|L\rangle)_b(\delta_1|b_1\rangle+\delta_2|b_2\rangle)$,
respectively.

Before the photon $a$ is put into the quantum circuit shown in
Fig.\ref{figure2}, a Hadamard operation is performed on its
spatial-mode with a 50:50 beam splitter (BS), which changes the
state of the photon $a$ to be
$|\phi_a\rangle_0=(\alpha_1|R\rangle+\alpha_2|L\rangle)_a(\gamma'_1|a_1\rangle+\gamma'_2|a_2\rangle)$.
Here $\gamma'_1=\frac{1}{\sqrt{2}}(\gamma_1+\gamma_2)$ and
$\gamma'_2=\frac{1}{\sqrt{2}}(\gamma_1-\gamma_2)$. The state of the
system composed of the QD and the photon $a$ is changed from
$|\phi_{ae}\rangle_0$ to be $|\phi_{ae}\rangle_1$ after the photon
$a$ passes through X$_1$, QD, X$_2$, CPBS, Z$_1$ and Z$_2$. Here
\begin{eqnarray}                           \label{eq.7}
|\phi_{ae}\rangle_0&=&\frac{1}{\sqrt{2}}(|\uparrow\rangle+|\downarrow\rangle)_e(\alpha_1|R\rangle+\alpha_2|L\rangle)_a(\gamma'_1|a_1\rangle+\gamma'_2|a_2\rangle),\nonumber\\
|\phi_{ae}\rangle_1&=&\frac{-1}{\sqrt{2}}\big[|\uparrow\rangle_e(\gamma'_2|a_1\rangle+\gamma'_1|a_2\rangle)
-|\downarrow\rangle_e(\gamma'_1|a_1\rangle
+\gamma'_2|a_2\rangle)\big](\alpha_1|R\rangle+\alpha_2|L\rangle)_a.\label{eq7}
\end{eqnarray}
It is not difficult to find that the outcome shown in Eq.(\ref{eq7})
is the result of the two-qubit CNOT gate using the electron spin $e$
and the spatial-mode of the photon $a$ as the control qubit and the
target qubit, respectively.

We perform a Hadamard operation on the electron spin $e$, and put
the photon $b$ into the quantum circuit shown in Fig.\ref{figure2}
with its initial state
$|\phi_b\rangle_0=(\beta_1|R\rangle+\beta_2|L\rangle)_b(\delta_1|b_1\rangle+\delta_2|b_2\rangle)$.
The state of the complete system composed of the QD, the photon $a$,
and the photon $b$ is changed from $|\varphi_{abe}\rangle_1$ to be
$|\varphi_{abe}\rangle_2$ after the photon $b$ passes through Z$_1$
and Z$_2$. Here the same operations are performed on the two photons
$a$ and $b$ by the QD-cavity system, X$_1$, X$_2$, CPBS, Z$_1$ and
Z$_2$, and the states $|\varphi_{abe}\rangle_1$ and
$|\varphi_{abe}\rangle_2$ are described as
\begin{eqnarray}                           \label{eq.8}
|\varphi_{abe}\rangle_1&\!\!=\!\!&|\phi_{ae}\rangle_1(\beta_1|R\rangle+\beta_2|L\rangle)_b(\delta_1|b_1\rangle+\delta_2|b_2\rangle),\nonumber\\
|\varphi_{abe}\rangle_2&\!\!=\!\!&\frac{1}{2}\big[\!-\!|\uparrow\rangle_e(\gamma'_1\!-\!\gamma'_2)(|a_1\rangle
\!-\!|a_2\rangle)(\delta_1|b_2\rangle \!+\! \delta_2|b_1\rangle)\nonumber\\
&&+|\downarrow\rangle_e(\gamma'_1 \!+\! \gamma'_2)(|a_1\rangle \!+\! |a_2\rangle)(\delta_1|b_1\rangle \!+\! \delta_2|b_2\rangle)\big]\nonumber\\
&& \otimes (\alpha_1|R\rangle \!+\!
\alpha_2|L\rangle)_a(\beta_1|R\rangle \!+\! \beta_2|L\rangle)_b.
\end{eqnarray}
At last, we perform Hadamard operations on the spatial-mode of the
photon $a$ and the electron spin $e$ in sequence, and the state of
the complete system becomes
\begin{eqnarray}                           \label{eq.9}
|\varphi_{abe}\rangle_3&\!=\!\!&\frac{1}{\sqrt{2}}\big\{|\uparrow\rangle_e\big[\gamma_1|a_1\rangle(\delta_1|b_1\rangle+\delta_2|b_2\rangle)
 -\gamma_2|a_2\rangle
(\delta_1|b_2\rangle+\delta_2|b_1\rangle)\big]-|\downarrow\rangle_e\big[\gamma_1|a_1\rangle(\delta_1|b_1\rangle\nonumber\\
&&+\delta_2|b_2\rangle)
 +\gamma_2|a_2\rangle(\delta_1|b_2\rangle+\delta_2|b_1\rangle)\big]\big\} (\alpha_1|R\rangle+\alpha_2|L\rangle)_a(\beta_1|R\rangle+\beta_2|L\rangle)_b.
\end{eqnarray}
The result of spatial-CNOT gate can be achieved by measuring the
state of the excess electron spin in the orthogonal basis
$\{|\uparrow\rangle,|\downarrow\rangle\}$ and making a feed-forward
operation. In detail, if the excess electron spin is in the state
$|\uparrow\rangle_e$, an additional sign change
$|a_2\rangle\rightarrow-|a_2\rangle$ is performed, and the result of
spatial-CNOT gate is obtained as
\begin{eqnarray}                           \label{eq.10}
|\psi_{ab}\rangle&\!\!=\!\!&\frac{1}{2}\big[\gamma_1|a_1\rangle(\delta_1|b_1\rangle
\!+\!\delta_2|b_2\rangle)
\!+\!\gamma_2|a_2\rangle(\delta_1|b_2\rangle
+\delta_2|b_1\rangle)\big](\alpha_1|R\rangle \!+\!
\alpha_2|L\rangle)_a(\beta_1|R\rangle \!+\!
\beta_2|L\rangle)_b.\;\;\;\;\;\;\;\;
\end{eqnarray}
\\

\begin{widetext}
\begin{center}
\begin{figure}[!h]
\centering
\includegraphics[width=12.8 cm,angle=0]{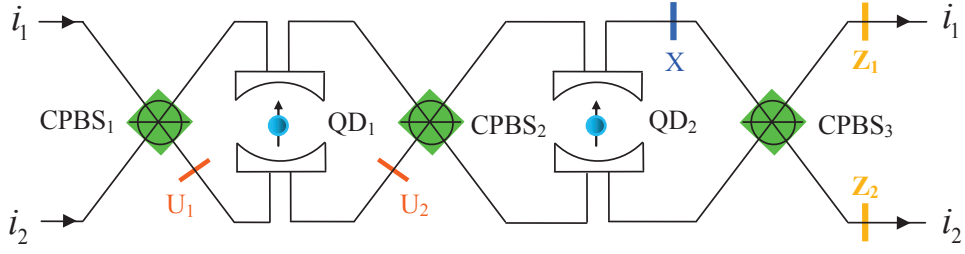}
\caption{Schematic diagram for a spatial-polarization hyper-CNOT
gate operating on both the spatial-mode and the polarization DOFs of
a two-photon system. $i_1$ and $i_2$ represent the two spatial modes
of photon $a$ or $b$.  U$_i$ ($i=1,2$) represents a wave plate which
is used to perform a polarization phase-flip operation
$U=-|R\rangle\langle R|-|L\rangle\langle L|$ on a photon.
\label{figure3}}
\end{figure}
\end{center}
\end{widetext}

\bigskip

\textbf{Spatial-polarization hyper-CNOT gate on a two-photon
system.}  We are going to discuss the construction of a
spatial-polarization hyper-CNOT gate which performs  CNOT gate
operations on both the spatial-mode and the polarization DOFs of a
two-photon system without using the auxiliary spatial modes or
polarization modes.

We have two QD-cavity systems to operate on the two DOFs of the
two-photon system $ab$, as shown in Fig.\ref{figure3}. Suppose the
two electron spins $e_1$ and $e_2$ in the two QD-cavity systems
(i.e., labeled as QD$_1$ and QD$_2$ in Fig.\ref{figure3}) are
prepared in the states
$\frac{1}{\sqrt{2}}(|\uparrow\rangle+|\downarrow\rangle)_{e_1}$ and
$\frac{1}{\sqrt{2}}(|\uparrow\rangle+|\downarrow\rangle)_{e_2}$,
respectively. The two photons $a$ and $b$ are prepared in the states
$|\psi_a\rangle_0=(\alpha_1|R\rangle+\alpha_2|L\rangle)_a(\gamma_1|a_1\rangle+\gamma_2|a_2\rangle)$
and
$|\psi_b\rangle_0=(\beta_1|R\rangle+\beta_2|L\rangle)_b(\delta_1|b_1\rangle+\delta_2|b_2\rangle)$,
respectively. We first perform Hadamard operations on both the
spatial-mode (with a 50:50 BS, not shown in Fig.\ref{figure3} and
the polarization (with two half-wave plates, not shown in
Fig.\ref{figure3}) DOFs of the photon $a$ before we put it into the
quantum circuit shown in Fig.\ref{figure3}, and the state of photon
$a$ is changed to be
$|\psi'_{a}\rangle_0=(\alpha'_1|R\rangle+\alpha'_2|L\rangle)_a(\gamma'_1|a_1\rangle+\gamma'_2|a_2\rangle)$.
Here $\alpha'_1=\frac{1}{\sqrt{2}}(\alpha_1+\alpha_2)$,
$\alpha'_2=\frac{1}{\sqrt{2}}(\alpha_1-\alpha_2)$,
$\gamma'_1=\frac{1}{\sqrt{2}}(\gamma_1+\gamma_2)$, and
$\gamma'_2=\frac{1}{\sqrt{2}}(\gamma_1-\gamma_2)$. Subsequently, we
let the photon $a$ pass through CPBS$_1$, U$_1$, QD$_1$, U$_2$, and
CPBS$_2$ in sequence, and the state of the system composed of QD$_1$
and the photon $a$ is changed from $|\Psi_{ae_1}\rangle_0$ to be
$|\Psi_{ae_1}\rangle_1$. Here
\begin{eqnarray}                           \label{eq.11}
|\Psi_{ae_1}\rangle_0&\!\!=\!\!&\frac{1}{\sqrt{2}}(|\uparrow\rangle+|\downarrow\rangle)_{e_1}(\alpha'_1|R\rangle+\alpha'_2|L\rangle)_a (\gamma'_1|a_1\rangle+\gamma'_2|a_2\rangle),\nonumber\\
|\Psi_{ae_1}\rangle_1&\!\!=\!\!&\frac{1}{\sqrt{2}}\{\gamma'_1\big[|\uparrow\rangle_{e_1}(\alpha'_1|L\rangle+\alpha'_2|R\rangle)_a
 +|\downarrow\rangle_{e_1}(\alpha'_1|R\rangle+\alpha'_2|L\rangle)_a)\big]|a_2\rangle
\nonumber\\
&&+\gamma'_2\big[|\uparrow\rangle_{e_1}(\alpha'_1|R\rangle+\alpha'_2|L\rangle)_a
+|\downarrow\rangle_{e_1}(\alpha'_1|L\rangle+\alpha'_2|R\rangle)_a)\big]|a_1\rangle\}.\label{eq10}
\end{eqnarray}
We let the photon $a$ pass through QD$_2$, X, CPBS$_3$, Z$_1$, and
Z$_2$ in sequence, and the state of the system composed of the
photon $a$ and the two electron spins in QD$_1$ and QD$_2$ is
changed from $|\Phi_{ae_1e_2}\rangle_1$ to be
$|\Phi_{ae_1e_2}\rangle_2$. Here
\begin{eqnarray}                           \label{eq.12}
|\Phi_{ae_1e_2}\rangle_1&\!\!=\!\!&\frac{1}{\sqrt{2}}(|\uparrow\rangle+|\downarrow\rangle)_{e_2}|\Psi_{ae_1}\rangle_1,\nonumber\\
|\Phi_{ae_1e_2}\rangle_2&\!\!=\!\!&\frac{1}{2}\big[|\uparrow\rangle_{e_1}(\alpha'_1|R\rangle+\alpha'_2|L\rangle)_a
 +|\downarrow\rangle_{e_1}(\alpha'_2|R\rangle+\alpha'_1|L\rangle)_a)\big]
\nonumber\\
&&
\otimes\big[|\uparrow\rangle_{e_2}(\gamma'_2|a_1\rangle+\gamma'_1|a_2\rangle)
 -|\downarrow\rangle_{e_2}(\gamma'_1|a_1\rangle+\gamma'_2|a_2\rangle)\big].\label{eq12}
\end{eqnarray}
This is the result of a four-qubit CNOT gate using the electron
spins $e_1$ and $e_2$ as the control qubits and the polarization and
the spatial-mode DOFs of the photon $a$ as the target qubits,
respectively, and the two CNOT gate operations are performed
independently without disturbing the state of photon $a$ in the
other DOF.

Subsequently, we let the photon $b$ pass through the quantum circuit
shown in Fig.\ref{figure3} after Hadamard operations are performed
on the electron spins $e_1$ in QD$_1$ and $e_2$ in QD$_2$. The state
of the system composed of the two photons and the two electron spins
is changed from $|\Xi_{abe_1e_2}\rangle_2$ to be
$|\Xi_{abe_1e_2}\rangle_3$ after the photon $b$ passes through
CPBS$_1$, U$_1$, QD$_1$, U$_2$, CPBS$_2$, QD$_2$, X, CPBS$_3$,
Z$_1$, and Z$_2$. Here the two states $|\Xi_{abe_1e_2}\rangle_2$ and
$|\Xi_{abe_1e_2}\rangle_3$ are described as
\begin{eqnarray}                           \label{eq.13}
|\Xi_{abe_1e_2}\rangle_2&\!\!=\!\!&|\Phi_{ae_1e_2}\rangle_2(\beta_1|R\rangle+\beta_2|L\rangle)_b (\delta_1|b_1\rangle+\delta_2|b_2\rangle),\nonumber\\
|\Xi_{abe_1e_2}\rangle_3&\!\!=\!\!&\frac{1}{2}\big[|\uparrow\rangle_{e_1}\alpha_1(|R\rangle+|L\rangle)_a(\beta_1|R\rangle+\beta_2|L\rangle)_b\nonumber\\
&&+|\downarrow\rangle_{e_1}\alpha_2(|R\rangle-|L\rangle)_a(\beta_2|R\rangle+\beta_1|L\rangle)_b\big]\nonumber\\
&&\otimes \big[-|\uparrow\rangle_{e_2}\gamma_2(|a_1\rangle-|a_2\rangle)(\delta_2|b_1\rangle+\delta_1|b_2\rangle)\nonumber\\
&&+|\downarrow\rangle_{e_2}\gamma_1(|a_1\rangle+|a_2\rangle)(\delta_1|b_1\rangle+\delta_2|b_2\rangle)\big].
\label{eq13}
\end{eqnarray}

At last, we perform Hadamard operations on the spatial-mode and the
polarization DOFs of  the  photon $a$ and the excess electron spins
in QD$_1$ and QD$_2$ again, and the state of the photon-electron
spin system is changed to be
\begin{eqnarray}                           \label{eq.14}
|\Xi_{abe_1e_2}\rangle_4&\!\!=\!\!&\frac{1}{2}\big\{|\uparrow\rangle_{e_1}\big[\alpha_1|R\rangle_a(\beta_1|R\rangle+\beta_2|L\rangle)_b
  +\alpha_2|L\rangle_a(\beta_2|R\rangle+\beta_1|L\rangle)_b\big]
\nonumber\\
&&
+|\downarrow\rangle_{e_1}\big[\alpha_1|R\rangle_a(\beta_1|R\rangle+\beta_2|L\rangle)_b -\alpha_2|L\rangle_a(\beta_2|R\rangle+\beta_1|L\rangle)_b\big]\big\}\nonumber\\
&& \otimes
\big\{|\uparrow\rangle_{e_2}\big[\gamma_1|a_1\rangle(\delta_1|b_1\rangle+\delta_2|b_2\rangle)
-\gamma_2|a_2\rangle(\delta_2|b_1\rangle+\delta_1|b_2\rangle)\big]
\nonumber\\
&&-|\downarrow\rangle_{e_2}\big[\gamma_1|a_1\rangle(\delta_1|b_1\rangle+\delta_2|b_2\rangle)
+\gamma_2|a_2\rangle(\delta_2|b_1\rangle+\delta_1|b_2\rangle)\big]\big\}.\label{eq15}
\end{eqnarray}

By measuring the two excess electron spins in the orthogonal basis
$\{|\!\!\uparrow\rangle,|\!\!\downarrow\rangle\}$ and performing an
additional sign change $|L\rangle_a\rightarrow-|L\rangle_a$ when the
electron spin $e_1$ is in the state $|\downarrow\rangle_{e_1}$ and
an additional sign change $|a_2\rangle\rightarrow-|a_2\rangle$ when
the electron spin $e_2$ is in the state $|\uparrow\rangle_{e_2}$,
the state of the two-photon system $ab$ becomes
\begin{eqnarray}                           \label{eq.15}
|\psi_{ab}\rangle&\!\!=\!\!&
\big[\alpha_1|R\rangle_a(\beta_1|R\rangle+\beta_2|L\rangle)_b
+\alpha_2|L\rangle_a(\beta_2|R\rangle+\beta_1|L\rangle)_b \big]
\nonumber\\
&&\otimes
 \big[\gamma_1|a_1\rangle(\delta_1|b_1\rangle+\delta_2|b_2\rangle) +\gamma_2|a_2\rangle(\delta_2|b_1\rangle+\delta_1|b_2\rangle)
\big].\label{eq16}
\end{eqnarray}
It is not difficult to find that there is a bit flip on the
spatial-mode of the photon $b$ (the target qubit) when the
spatial-mode of the photon $a$ (the control qubit) is $\vert
a_2\rangle$, compared with the initial state of the two-photon
system $ab$. Moreover, a bit flip takes place on the polarization of
the photon $b$ when the polarization of  the  photon $a$ is $\vert
L\rangle$. That is, the quantum circuit shown in Fig.\ref{figure3}
can be used to achieve a hyper-CNOT gate operating on the two-photon
system $ab$ on both its polarization and its spatial-mode DOFs
simultaneously, without auxiliary spatial modes or polarization
modes. Moreover, the success probability of this hyper-CNOT gate is
100\% in principle.


\bigskip

{\large \textbf{Discussion}}

In experiment, the transmission and the reflection rules in
Eqs.(\ref{eq5}) and (\ref{eq6}) may fail because of decoherence and
dephasing. In this time, the fidelity and the efficiency of our
hyper-CNOT gate are reduced. The spin-dependent transition rule is
not perfect if we consider the heavy-light hole mixing which can
reduce the fidelity by a few percent, while this hole mixing can be
reduced by improving the shape, size, and type of QDs\cite{QD4}. The
fidelity can be also reduced in only a few percent by the exciton
dephasing, including optical dephasing and hole spin dephasing. The
optical coherence time of exciton ($\sim100p$s) is ten times longer
than the cavity photon lifetime\cite{trion1,trion2}, and the hole
spin coherence time is three times order of magnitude longer than
the cavity photon lifetime\cite{trion5,trion6}. The fine structure
splitting, which occurs for neutral exciton, is immune for charged
exciton due to the quenched exchange interaction\cite{fine,fine1}.
The electron spin decoherence is another factor which can reduce the
fidelity, but it can be reduced by extending electron coherence time
to $\mu$s using spin echo techniques, which is longer than the
cavity photon lifetime ($\sim10p$s)\cite{QD4}. The preparation of
spin superposition states $|+\rangle$ and $|-\rangle$ can be
implemented by using nanosecond electron spin resonance microwave
pulses or picosecond optical pulses\cite{manu}, of which the
preparation time ($p$s) is significantly shorter than spin coherence
time\cite{QD2,pluse}. Then the Hadamard operation for transforming
electron spin states $|\uparrow\rangle$ and $|\downarrow\rangle$ to
$|+\rangle$ and $|-\rangle$ can be achieved.

In the resonant condition with
$\omega_c=\omega_{X^-}=\omega_0=\omega$, if the cavity side leak is
taken into account, the optical selection rules for a QD--cavity
system given by Eqs. (\ref{eq5}) and (\ref{eq6}) become\cite{QD3}
\begin{eqnarray} \label{eq16}
|R^\uparrow, i_2, \uparrow\rangle      &\rightarrow&   |r||L^\downarrow, i_2, \uparrow\rangle   - |t||R^\uparrow, i_1, \uparrow\rangle,\nonumber\\
|L^\downarrow, i_1,  \uparrow\rangle   &\rightarrow&   |r||R^\uparrow, i_1, \uparrow\rangle     - |t||L^\downarrow, i_2,  \uparrow\rangle,\nonumber\\
|R^\downarrow, i_1, \downarrow\rangle  &\rightarrow&   |r||L^\uparrow, i_1, \downarrow\rangle   - |t||R^\downarrow, i_2, \downarrow\rangle,\nonumber\\
|L^\uparrow , i_2,  \downarrow\rangle  &\rightarrow&   |r||R^\downarrow, i_2, \downarrow\rangle - |t||L^\uparrow , i_1, \downarrow\rangle,\nonumber\\
|R^\downarrow, i_1, \uparrow\rangle    &\rightarrow&  -|t_0||R^\downarrow, i_2, \uparrow\rangle  + |r_0||L^\uparrow, i_1, \uparrow\rangle,\nonumber\\
|L^\uparrow , i_2,  \uparrow\rangle    &\rightarrow&  -|t_0||L^\uparrow , i_1,  \uparrow\rangle  + |r_0||R^\downarrow, i_2,  \uparrow\rangle,\nonumber\\
|R^\uparrow, i_2, \downarrow\rangle    &\rightarrow&  -|t_0||R^\uparrow, i_1, \downarrow\rangle  + |r_0||L^\downarrow, i_2, \downarrow\rangle,\nonumber\\
|L^\downarrow, i_1, \downarrow\rangle  &\rightarrow&
-|t_0||L^\downarrow, i_2, \downarrow\rangle+|r_0||R^\uparrow, i_1,
\downarrow\rangle.
\end{eqnarray}

\begin{figure}[htb]                    
\centering
\includegraphics[width=6.2 cm]{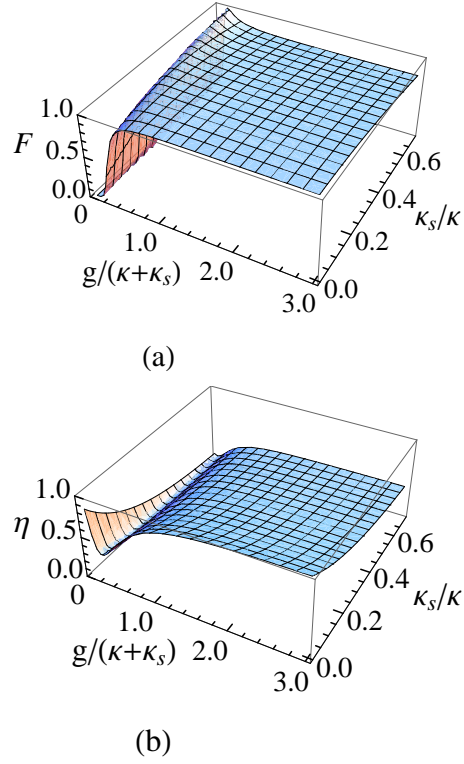}
\caption{Fidelity and efficiency of the present spatial-polarization
hyper-CNOT gate vs the coupling strength and side leakage rate with
$\gamma=0.1\kappa$. \label{figure4}}
\end{figure}

The fidelity of a quantum logical gate is defined as
$F=|\langle\psi_f|\psi\rangle|^2$, where $|\psi\rangle$ is the ideal
final state of the quantum system after the process for a quantum
logical gate, and $|\psi_f\rangle$ is the final state of the system
by considering external reservoirs. The efficiency of a photonic
logical gate is defined as the probability of the photons to be
detected after the logic gate operation. The fidelity and the
efficiency of  the present spatial-polarization hyper-CNOT gate are
\begin{eqnarray}                           \label{eq.17}
F&=&\frac{(\xi_1 \zeta_1 m)^2}{4(|\xi_1\zeta_2|^2+|\xi_1\xi_2|^2)n_1+4(|\xi_1\zeta_1|^2+|\xi_2^2|^2)n_2},\nonumber\\
\eta&=&\frac{1}{16}(|r|^2+|t|^2+|t_0|^2+|r_0|^2)^4.
\end{eqnarray}
Here
\begin{eqnarray}                           \label{eq.18}
\zeta_1&=&|t_0|+|r|-|r_0|-|t|,  \;\;\;\;\;\;\;\;\; \zeta_2=|t|-|r|-|r_0|+|t_0|, \nonumber\\
\xi_1&=&|t_0|+|r_0|+|r|+|t|, \;\;\;\;\;\;\;\;\; \xi_2=|r|+|t|-|r_0|-|t_0|,\nonumber\\
m&=&(|t_0|+|r_0|)|t_0|-(|r|+|t|)|r_0|+(|t_0|+|r_0|)|r|\nonumber\\
&&-(|r|+|t|)|t|-(|t_0|+|r_0|)|r_0|+(|r|+|t|)|t_0|\nonumber\\
&&-(|t_0|+|r_0|)|t|+(|r|+|t|)|r|,\nonumber\\n_1&=&|(|t|+|r|)|t_0|-(|r_0|+|t_0|)|r_0||^2+|(|t|+|r|)|r|-\nonumber\\
&&(|r_0|+|t_0|)|t||^2+|(|t|+|r|)|r_0|-(|r_0|+|t_0|)|t_0||^2\nonumber\\
&&+|(|t|+|r|)|t|-(|r_0|+|t_0|)|r||^2,\nonumber\\
n_2&=&|(|t_0|+|r_0|)|t_0|-(|r|+|t|)|r_0||^2+|(|t_0|+|r_0|)|r|\nonumber\\
&&-(|r|+|t|)|t||^2+|(|t_0|+|r_0|)|r_0|-(|r|+|t|)|t_0||^2\nonumber\\
&&+|(|t_0|+|r_0|)|t|-(|r|+|t|)|r||^2.
\end{eqnarray}
The fidelity and the efficiency are mainly affected by the coupling
strength and the cavity side leakage as shown in Fig.4\ref{figure4}.
The strong coupling strength and the low side leakage and cavity
loss rate ($\kappa_s/\kappa$) are required for the present
spatial-polarization hyper-CNOT quantum gate with a high fidelity
and a high efficiency. The strong coupling strength
$g/(\kappa+\kappa_s)\simeq0.8$ has been observed in a large
micropillar ($d=7.3$ $\mu$m)\cite{couple3} with a quality factor of
$Q\sim6.5\times10^4$ in a current experiment. In $d=1.5$ $\mu$m
micropillar  microcavity, the coupling strength
$g/(\kappa+\kappa_s)\simeq0.5$ was reported\cite{couple} with
$Q\sim8800$, and the coupling strength
$g/(\kappa+\kappa_s)\simeq2.4$ ($Q\sim4\times10^4$) could be
achieved\cite{couple1} by improving the sample designs, growth, and
fabrication\cite{couple2}. In the case
$g/(\kappa+\kappa_s)\simeq0.5$ and $\kappa_s/\kappa\simeq0$, the
fidelity and the efficiency for this hyper-CNOT gate are $F=92.5\%$
and $\eta=55\%$, respectively. If the coupling strength is
$g/(\kappa+\kappa_s)\simeq2.4$ with $\kappa_s/\kappa\simeq0$, the
fidelity and the efficiency  become $F=100\%$ and $\eta=97\%$,
respectively. If the side leakage and cavity loss rate is
$\kappa_s/\kappa\simeq0.3$ for $g/(\kappa+\kappa_s)\simeq2.4$, the
fidelity and the efficiency are $F=96\%$ and $\eta=60.3\%$. The
fidelity and the efficiency of this hyper-CNOT gate are mainly
reduced by a weak coupling strength and a high cavity side leakage.
The quality factor of a micropillar microcavity is dominated by the
side leakage and cavity loss rate $\kappa_s/\kappa$. The side
leakage and cavity loss rate has been reduced to
$\kappa_s/\kappa\simeq0.7$ with the quality factor reduced to
$Q\sim1.7\times10^4$ ($g/(\kappa+\kappa_s)\simeq1$) in the
micropillar $d=1.5$ $\mu$m by thinning down the top
mirrors\cite{QD4}. High efficiency operations are required to
achieve a stronger coupling strength with  a lower side leakage in
micropillars.

In our previous work, we introduced a hyper-CNOT gate constructed
with the reflection geometry optical property of one-side QD-cavity
systems with one mirror partially reflective and another mirror
100\% reflective\cite{HCNOT}. The hyper-CNOT gate in one-side
QD-cavity protocol is fragile because the reflectance for the
uncoupled cavity and the coupled cavity should be balanced to get
high fidelity\cite{QD3}. The present hyper-CNOT gate is constructed
with the reflection/transmission geometry optical property of
double-sided QD-cavity systems with both mirrors partially
reflective. And this hyper-CNOT gate is robust and flexible with the
large reflectance and transmittance difference between the uncoupled
cavity and the coupled cavity\cite{QD3}. Moreover, in the one-side
QD-cavity protocol, the problem about how to put the several spatial
modes of a photon into the single-sided cavity is required to be
solved. However, this is not a problem for this double-sided
QD-cavity protocol, because there are two balanced spatial modes for
a double-sided cavity. And the operations on the two spatial modes
are combined in this double-sided QD-cavity protocol, especially for
the Mach-Zehnder interferometer induced by the CPBSs. That is, both
the fidelity and the efficiency of the present hyper-CNOT gate are
higher than the one in one-side QD-cavity protocol.

With the hyper-CNOT gate operation, the quantum information
processing can be implemented with less photon resources, which can
reduce the resources required in the quantum logic circuits and the
storage process. And the photonic dissipation and the environment
noise effect are depressed in the quantum information processing
based on two DOFs of photon systems. Moreover, the interaction times
between the photons is decreased in the quantum information
protocols by using the two DOFs of photon systems, compared with the
integration of several cascaded CNOT gates in one DOF. For instance,
in the preparation of four-qubit cluster states, only a hyper-CNOT
gate operation (photons interact with electron spins four times) and
a wave plate are required in the protocol with two photons in two
DOFs\cite{HCNOT,HECP}, while three CNOT gate operations (photons
interact with electron spins six times) are required in the protocol
with four photons in one DOF. In this article, we have only
discussed the construction of a spatial-polarization hyper-CNOT gate
with the nonlinear optics of double-sided QD-cavity system. With the
present hyper-CNOT gate and single-photon quantum gates, a
hyper-parallel photonic quantum computation can be constructed in
principle, and the preparation and analysis of multi-photon
hyperentangled states can also be implemented in a simple way. In
principle, some other quantum systems with nonlinear optics can also
be used to achieve scalable hyper-parallel quantum computation, such
as wave-guide nanocavity, nitrogen-vacancy centers, Rydberg atoms,
cross-Kerr media, and so on. Here the hyper-CNOT gate operation is
used to operate on the polarization and the spatial-mode DOFs
independently, and the two photons with two DOFs are used as four
qubits. In fact, the two photons with two DOFs
 can also be used as two qudits in quantum information processing.
Moreover, as the CNOT gate on the two DOFs of a photon can be
implemented easily with the linear optical elements and the
single-qubit gates for the two DOFs of a photon can be implemented
independently, the multiqubit hybrid logic gate can be implemented
in a simpler way  with less photon resources resorting to the two
DOFs of the photon system.

In summary, we have constructed the spatial-polarization hyper-CNOT
gate operating on both the spatial-mode and the polarization DOFs of
a two-photon system with the giant optical circular birefringence of
quantum dot inside double-sided optical microcavity. This
spatial-polarization hyper-CNOT gate may work efficiently in strong
coupling regimes with low cavity  side leakage. In the construction
of the hyper-CNOT gate, the two DOFs of a two-photon system are used
to assist each other to implement the CNOT gate operations without
using the auxiliary spatial modes, which is much simpler than the
one constructed with the quantum dots inside one-side optical
microcavities\cite{HCNOT}. The measurement on the electron spins in
the QDs can be replaced by making the control qubits of the photon
system interact  with QD-cavity systems twice as that introduced by
Duan and Kimble\cite{QED1}, but this will decrease the fidelity and
the efficiency of the hyper-CNOT gate. Therefore, it relies on the
experimental technique for deciding whether people use the
measurement on electron spins or not.

\bigskip

{\large \textbf{Methods}}

\textbf{Measurement on electron spin $e$ in QD.}  The measurement on
the electron spin $e$ can be implemented by using an auxiliary
photon. If the auxiliary photon is initially in the state
$|\psi_i\rangle=|R_i\rangle|i_1\rangle$, after interacting with
QD-cavity system, the final state of the complete system becomes
\begin{eqnarray}                           \label{eq.19}
|R_i\rangle|i_1\rangle|\uparrow\rangle_e&\rightarrow&-|R_i\rangle|i_2\rangle|\uparrow\rangle_e,\;\;\;\;\;\;\;\;\nonumber\\
|R_i\rangle|i_1\rangle|\downarrow\rangle_e&\rightarrow&|L_i\rangle|i_1\rangle|\downarrow\rangle_e.
\end{eqnarray}
If the auxiliary photon is detected on the spatial mode
$|i_2\rangle$, the electron spin is projected into the state
$|\uparrow\rangle_e$, whereas the electron spin is projected into
the state $|\downarrow\rangle_e$ with the detection of the auxiliary
photon on the spatial mode $|i_1\rangle$.

\bigskip

\textbf{Spin echo technique using single photon pulse.} Our schemes,
implemented with the optical properties of QD-cavity system, are not
compatible with ESR-based spin echo technique in an external
magnetic field, which may cause the detuning between the two spin
states. In 2011, Hu and Rarity\cite{QD4} showed that the microwave
pulses or optical pulses used in spin echo techniques may be
replaced by the single photon pulses to suppress the nuclear spin
fluctuations and hyperfine interaction. That is, the single photons
can play the role of the $\pi-$pulse to make the spin rotate
$180^\circ$ around the optical axis. If a photon in the state
$\frac{1}{\sqrt{2}}(|R\rangle+|L\rangle)_i|i_1\rangle$ is put in to
the CPBS, QD and CPBS in sequence, after interaction, the state of
the system composed of the photon $i$ and the electron spin $e$ in
QD is changed as
\begin{eqnarray}                           \label{eq.20}
\frac{1}{2}(|R\rangle+|L\rangle)_i|i_1\rangle(|\uparrow\rangle+|\downarrow\rangle)_e\;\;\rightarrow
\;\;
\frac{1}{2}(|R\rangle+|L\rangle)_i|i_2\rangle(|\uparrow\rangle-|\downarrow\rangle)_e.
\end{eqnarray}

\bigskip

\textbf{Double-sided QD-cavity system in quantum information
processing based on two DOFs.}  As the spatial-mode and the
polarization DOFs of a photon system can be converted into each
other  easily, the logic gate operation on the polarization
(spatial-mode) DOF may be simplified by assisting with the
spatial-mode (polarization) DOF, especially for the interaction
between the two photons. There are many works on the polarization
logic gates using the spatial-mode DOF as the
assistant\cite{linear2,nonlinear,QED1,QD6,QD5,QDWei1,QD7}. The
double-sided QD-cavity system has two spatial modes, which are
always used as assistant in manipulating the polarization DOF and at
last be consumed\cite{QD6,QD5,QDWei1,QD7}. The spatial-mode and the
polarization DOFs of a photon can be manipulated independently and
easily with linear optical elements, so it is convenient to use
these two DOFs as two qubits in quantum information processing. Our
hyper-CNOT gate shows that the spatial-mode and the polarization
DOFs can assist each other to perform CNOT gate operations on the
photonic states of these two DOFs independently without auxiliary
spatial modes, resorting to the optical property shown in
Eqs.(\ref{eq5}) and (\ref{eq6}). This is simpler than the one using
one-side QD-cavity system\cite{HCNOT}. Of course, the two DOFs of a
photon can be used as a qudit, and the double-sided QD-cavity system
can also be used to implement the quantum information processing
with multiple qudits in two DOFs. It may be convenient to use
double-sided QD-cavity system to implement the quantum information
processing with two DOFs.

\bigskip
\bigskip

{\large \textbf{Acknowledgments}}

This work was supported by the National Natural Science Foundation
of China under Grant No. 11174039 and  NECT-11-0031.

\bigskip

{\large \textbf{Authors contributions}}

B.C. and F.G.  wrote  the main  manuscript text  and  B.C. prepared
figures 1-4. Both authors reviewed the manuscript.

\bigskip

{\large \textbf{Additional information}}

Competing financial interests: The authors declare no competing
financial interests.

\end{document}